\documentclass[letterpaper,english,prb,superscriptaddress,twocolumn]{revtex4}

\usepackage{graphics}

\makeatletter

\providecommand{\LyX}{L\kern-.1667em\lower.25em\hbox{Y}\kern-.125emX\@}

\makeatother

\begin{document}

\title{Magnetic excitations of spin and orbital moments in cobalt oxide }

\author{Z. Yamani}\email{Zahra.Yamani@nrc.gc.ca}\affiliation{National Research Council, Canadian Neutron Beam Centre, Chalk River, ON K0J 1J0, Canada}
\author{W.J.L. Buyers}
\affiliation{National Research Council, Canadian Neutron Beam Centre, Chalk River, ON K0J 1J0, Canada} \affiliation{Canadian Institute of Advanced Research,
Toronto, Ontario, Canada M5G 1Z8}
\author{R.A. Cowley}
\affiliation{Clarendon Laboratory, University of Oxford, Parks Road, Oxford OX1 3PU, UK}
\author{D. Prabhakaran}
\affiliation{Clarendon Laboratory, University of Oxford, Parks Road, Oxford OX1 3PU, UK}

\begin{abstract}

Magnetic and phonon excitations in the antiferromagnet CoO with an unquenched orbital angular momentum are studied by neutron scattering. Results of energy scans in several Brillouin zones in the (HHL) plane for energy transfers up to 16 THz are presented. The measurements were performed in the antiferromagnetic ordered state at 6 K (well below T$_N$$\sim$290 K) as well as in the paramagnetic state at 450 K. Several magnetic excitation modes are identified from the dependence of their intensity on wavevector and temperature. Within a Hund's rule model the excitations correspond to fluctuations of coupled orbital and spin degrees of freedom whose bandwidth is controlled by interionic superexchange. The different $<$111$>$ ordering domains give rise to several magnetic peaks at each wavevector transfer.

\end{abstract}
\maketitle

\section{Introduction}\label{labelOfFirstSection}

Transition metal oxides exhibit a wide range of behaviour including colossal magnetoresistance, orbital and charge order, superconductivity and formation of high spin or low-spin states~\cite{TMObook}. These are all related to their magnetic properties. Surprisingly, even in simple rock salt structure compounds an accepted account of their magnetism is still lacking. The theory is least certain and most challenging for compounds for which the atomic orbital moment as well as the spin moment contribute to their magnetic properties. The magnetic ground state is split by spin-orbit interaction, as well as by a spin exchange field that can mix different orbital states and so cause a large spin gap in the magnetic excitation spectrum.

Among transition metal oxides, oxides containing cobalt have recently attracted considerable attention~\cite{boothroyd04,koshibae03}. This is because the Co ion has several different valence states (i.e. Co$^{2+}$ in CoO, Co$^{3+}$ in LiCoO$_2$, and Co$^{4+}$ in CoO$_2$) as well as transitions between high and low spin states giving rise to charge ordering, magnetic ordering and superconductivity as the crystal field and doping of the materials are changed. Neutron scattering is well suited to determining the spin-orbit coupling strength as well as the exchange. In one of the simplest oxides, CoO, however the electronic and magnetic ground state is poorly understood~\cite{sakurai68,yamani08,tomiyasu06,ressouche06,jauch04}.

CoO has a face-centered-cubic rock-salt type structure with Fm\={3}m space group in its high temperature paramagnetic phase. It orders antiferromagnetically (AF) below T$_N$$\sim$290 K with the type-II structure~\cite{roth58}. The transition is believed to have a weak first order nature~\cite{oleaga09}. Below the N\'{e}el transition, there is a change to a less symmetric crystal structure with a monoclinic unit cell. With a $\sim$1$\%$ contraction along a cube edge being the dominant distortion, the low temperature phase is usually approximated by a tetragonal structure.

In CoO the nearest neighbour antiferromagnetic exchange is frustrated, contributing no molecular field, and it is the antiferromagnetic next nearest neighbour exchange that breaks the symmetry below T$_N$. Magnetic ordering occurs with (111) ferromagnetic sheets of spins stacking antiferromagnetically along [111] directions. The ordered moment at low temperature has been measured~\cite{roth58} to be 3.98(6) Bohr magnetons. It is suggested~\cite{tomiyasu04} that the magnetic structure is the result of two wavevectors added together but the dominant one has a propagating wavevector of $\mathbf{Q}$=(0.5 0.5 0.5) and the direction of the moment is approximately (-0.325 -0.325 0.888). This corresponds to an angle of 27 degrees from the (0 0 1) axis and lies close to the (-1 -1 2) direction that is perpendicular to (0.5 0.5 0.5).

In the earliest measurement of the magnetic excitations of single crystalline CoO only two broad spin excitations were observed~\cite{sakurai68}. The early results were unexpected because the magnetic excitations were largely independent of temperature and consisted of flat branches with almost no dispersion. A mean field theory was developed that had two quite different explanations for the data taken at 330 K and 100 K even though the original data was very similar.

Instead the spin excitations should display a rich spectrum. A free Co$^{2+}$ ion has a 3$d^7$ electronic configuration i.e. three holes in the $d$-shell. According to Hund's rule, the ground state of this ion has a total spin $S$=3/2, and total angular momentum $L$=3. In a crystal field with cubic local symmetry, the lowest orbital state is a triplet described~\cite{buyers71} by an effective angular momentum $l$=1. The spin-orbit coupling, then, takes the form (-3/2)$\lambda_1 \textbf{l}~.~\textbf{S}$ where $\lambda_1$ is the free-ion spin-orbit parameter reduced by a factor due to covalency. Magnetic excitations correspond to coupled spin and orbital excitations because the orbital angular momentum is unquenched. In the internal spin exchange field the total angular momentum, $\textbf{j}=\textbf{l}+\textbf{S}$, is no longer a good quantum number and the states are split and mixed. Excitations where the orbital state has a large change are called spin orbitons. Hence the magnetic excitations are mixed and have both spin and orbital components. Identification of the different modes is rendered complex by the presence of three domains from the tetragonal compression along each of the cubic axes, causing crystal field splitting, and by the four ordering domains associated with the four $\{$111$\}$ directions of the antiferromagnetic sheets. Moreover the spin direction remains an issue.

In order to understand the magnetic excitations in CoO, we have undertaken detailed neutron scattering studies of a high quality CoO single crystal in both its ordered and paramagnetic states. Our results show that there are at least four resolved excitation peaks in the spectrum between 4 and 12 THz at 6 K. Comparison with the Co$^{2+}$ magnetic form factor shows that the three higher energy peaks have substantial magnetic weight while the lowest energy peak is only weakly magnetic. This is also supported by the observed temperature dependence of these peaks. Our observation of four sharp peaks at 6 K differs from the original work that reported~\cite{sakurai68} only two broad peaks at 110 K. Thus 110 K was not a sufficiently low temperature to observe the many sharp excitations seen at the low temperature limit. Our results are also an improvement on a recent lower energy resolution experiment where only two peaks were reported~\cite{tomiyasu06}.

\section{Experiment}

A large 10 g high quality single crystal was grown at Oxford. We performed inelastic neutron scattering measurements at the C5 spectrometer at the NRU reactor, Chalk River. The crystal was aligned in the (HHL) scattering plane. In the first set of experiments (PG-PG configuration), we used a vertically focusing PG(002) monochromator and flat PG analyzer with a fixed final energy E$_f$=3.52 THz with collimations selected to give an energy resolution of 0.3 THz at zero energy transfer and 1.2 THz at 10 THz energy. To improve resolution in the 6 to 12 THz energy range, we later used Be(002) as monochromator and PG(002) as analyzer (Be-PG configuration) with a fixed final energy E$_f$=7.37 THz and tighter collimations. The resolution was reduced to 0.8 THz at 10 THz energy transfer. With this configuration similar spectra to the PG-PG configuration were measured while the resolution was tight enough to be able to distinguish individual modes in the scattering in the range 6 to 12 THz. In both configurations higher order neutrons were removed by a graphite filter.

Since phonons may cause additional peaks and interact with magnetic excitations, we measured the phonon and magnetic response also at 450 K with the PG-PG configuration. There the magnetic excitations give only broad paramagnetic scattering and the phonons appear as well-defined peaks. High temperature was important in order to eliminate the average exchange field and to find the exchange-free energy of the first spin-orbit state so as to determine the spin-orbit coupling constant.

\section{Results and discussion}\label{results}

\begin{figure}[tbh!]
\begin{center}
\vskip 0cm
\resizebox{1.\linewidth}{!}{\includegraphics{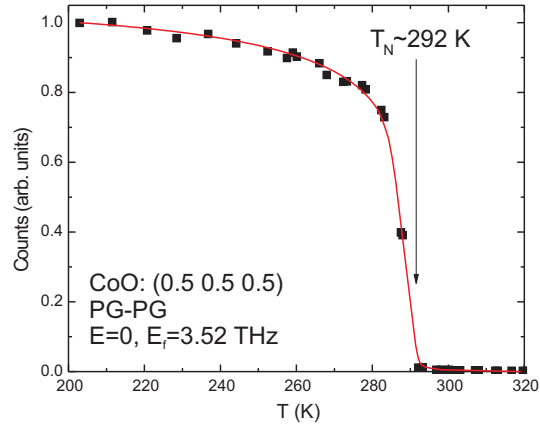}}\vskip 0cm \caption []{The temperature dependence of the order parameter is determined by measuring the temperature dependence of the elastic scattering at (0.5 0.5 0.5) wavevector. This measurement indicates the transition occurs at $\sim$292 K. The solid line is a guide to the eye.} \label{fig1}
\end{center}
\end{figure}

Fig.~\ref{fig1} shows the observed elastic scattering at \textbf{Q}=(0.5 0.5 0.5), the ordering wavevector as a function of temperature. The transition to type II AF order occurs at $\sim$292 K, similar to previous reports~\cite{roth58,sakurai68}.

\begin{figure}
\begin{center}
\vskip 0cm
\resizebox{1.05\linewidth}{!}{\includegraphics{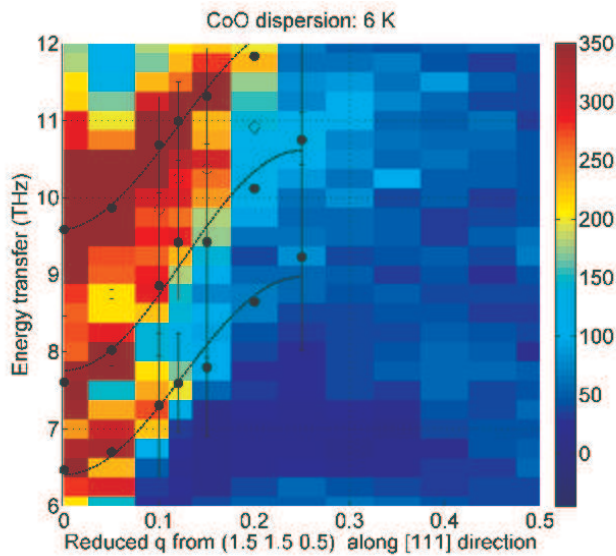}}\vskip 0cm \caption []{The magnetic excitations observed at 6 K in the radial direction from (1.5 1.5 0.5) for up to 12 THz. Symbols are from fits to the constant-q scans similar to Fig.~\ref{fig3}. The error bar associated with each data point is the FWHM of each peak from the fits.  } \label{fig2}
\end{center}
\end{figure}

\begin{figure} 
\begin{center}
\vskip 0cm
\resizebox{1\linewidth}{!}{\includegraphics{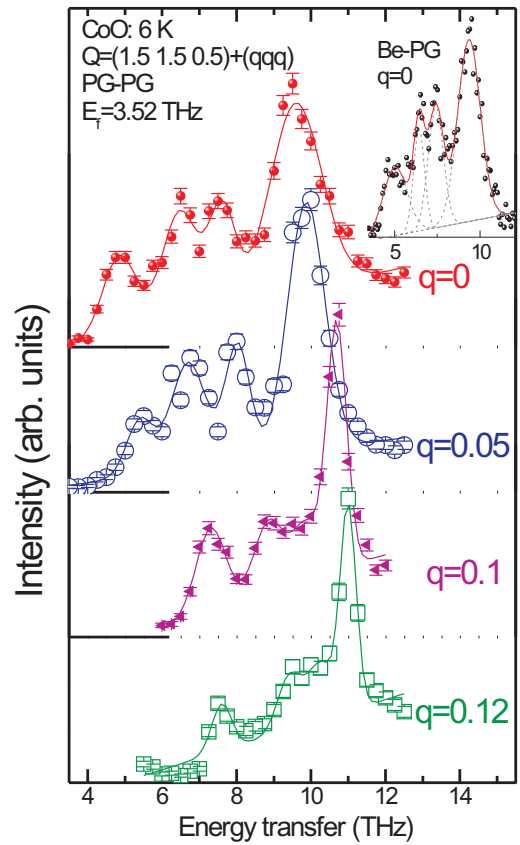}}\vskip 0cm \caption []{Constant-\textbf{Q} scans measured at 6 K at (1.5 1.5 0.5)+(q, q, q) in the (111) direction. The reduced q is noted for each scan. The inset shows the higher resolution spin spectrum measured with Be-PG configuration with E$_f$=7.37 THz at (1.5 1.5 0.5) at 12 K. There are at least four branches. Fits to these branches are shown with solid lines.} \label{fig3}
\end{center}
\end{figure}

The magnetic excitations at 6 K are shown in Fig.~\ref{fig2} in the frequency range 6 to 12 THz. This data is obtained from constant-\textbf{Q} scans similar to scans shown in Fig.~\ref{fig3} for reduced wavevectors, (q, q, q),  measured with respect to (1.5 1.5 0.5) in the (111) direction. These measurements indicate that there are at least four q=0 peaks between 4 and 12 THz at 4.9, 6.5, 7.6 and 9.5 THz. The presence of these peaks is confirmed with the high energy resolution Be-PG experiment shown in the inset of Fig.~\ref{fig3}. The zone-centre peak at 9.5 THz is the strongest peak and it is broader than both the resolution and the other peaks. This indicates that this peak can be the result of overlap of two or more peaks. In fact for higher q, this peak becomes narrower and it is possible to fit two peaks to this excitation, see the fits in Fig.~\ref{fig3}. At energies less than 4 THz there are no low energy modes of appreciable strength at any measured wavevector except for well defined phonons near the nuclear Bragg peaks. The peak at 4.9 THz exists throughout the Brillouin zone and we associate this peak in part with incoherent scattering. At small q, the peaks at 6.5, 7.6 and 9.5 THz exhibit strong upward dispersion as q increases. Measurements from other zones also show similar behaviour, all indicating that the strongest band of excitations rises from a zone-centre energy of 9.5 THz.

\begin{figure} 
\begin{center}
\vskip 0cm
\resizebox{1\linewidth}{!}{\includegraphics{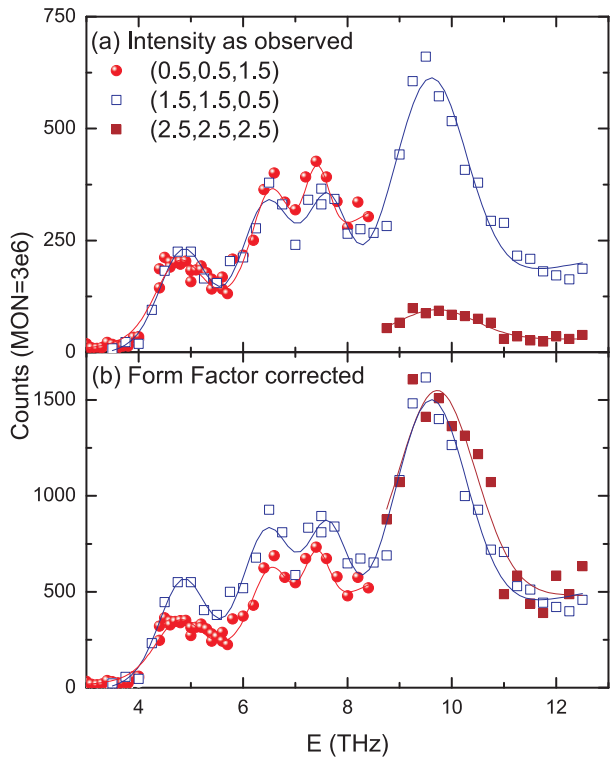}}\vskip 0cm \caption []{(a) The magnetic excitations at different zone centres as observed at 6 K. (b) The same data is shown after form factor correction. The form factor analysis indicates that the peak at 9.5 THz is mainly magnetic while the peaks at lower frequencies, 6.5 and 7.6 THz, have partial magnetic weight. The peak at 4.9 THz appears to have the least magnetic weight.} \label{fig4}
\end{center}
\end{figure}

A comparison between the observed intensities at different wavevectors with the expected values from the magnetic form factor can be used to identify magnetic from phonon excitations. We have calculated the Co$^{2+}$ magnetic form factor using the tabulated integrals~\cite{FFbrown06} for Co$^{2+}$ which are in good agreement with experimental data for CoO~\cite{mahendra71}. The squares of the magnetic form factors calculated at several zone centre wavevectors are listed in Table 1. Fig.~\ref{fig4}(a) shows the observed intensity at three zone centres (see Fig. 1 of Ref.~\cite{yamani08} for the data at a few more zone centres). The observed intensities divided by the magnetic form factor squared are shown in the Fig.~\ref{fig4}(b). This analysis indicates that the peak at 9.5 THz is mainly magnetic in origin. The intensities of the peaks at 6.5 and 7.6 THz decrease with increasing Q, but by less than the form factor, suggesting the peaks at 6.5 and 7.6 THz have partial magnetic weight. The peak at 4.9 THz is observed to have a similar cross-section at (0.5 0.5 1.5), (1.5 1.5 0.5) and (2.5 2.5 0.5) zone centres (see Fig.~\ref{fig4}(a) and Fig. 1 in Ref.~\cite{yamani08}). The coherent cross-section for TA phonons with \textbf{q}=$<$0.5 0.5 0.5$>$ also has the same value at these three wavevectors despite their different $|\textbf{Q}|$. Hence we believe that this lowest peak arises largely from coherent phonon scattering, as well as incoherent phonon scattering since it coincides with the possible peak in the density of phonon states~\cite{wdowik07}.

\begin{table}[ht]
\caption{Magnetic form factor squared for Co$^{2+}$ at several magnetic zone centres~\cite{mahendra71}. } % title of Table
\centering % used for centering table
\begin{tabular}{c c} % centered columns (2 columns)
\hline%\hline %inserts double horizontal lines
Zone Centre & Form Factor Squared \\ [0.5ex] % inserts table
%heading
\hline % inserts single horizontal line
  (0.5~0.5~0.5) & 0.86  \\
  (0.5~0.5~1.5) & 0.58  \\
  (1.5~1.5~0.5) & 0.41  \\
  (1.5~1.5~1.5) & 0.29  \\
  (2.5~2.5~0.5) & 0.13  \\
  (2.5~2.5~2.5) & 0.06  \\ [1ex] % [1ex] adds vertical space
\hline %inserts single line
\end{tabular}
\label{table:nonlin} % is used to refer this table in the text
\end{table}

\begin{figure} 
\begin{center}
\vskip 0cm
\resizebox{1\linewidth}{!}{\includegraphics{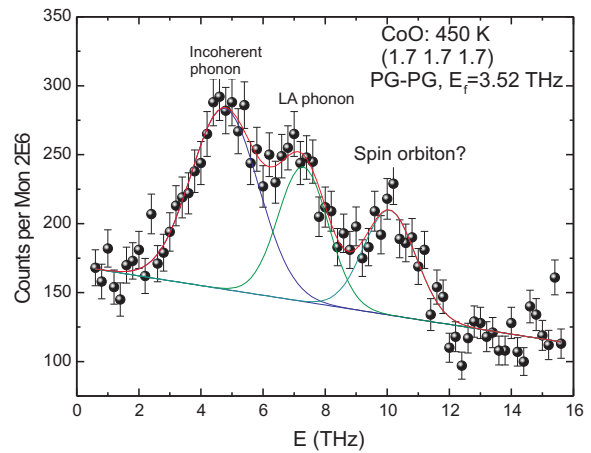}}\vskip 0cm \caption []{A scan at 450 K for \textbf{Q} = (1.7 1.7 1.7) showing the high-frequency orbiton transition at 10.2 THz, a LA phonon at 7.2 THz and an incoherent peak from cobalt vibrations at 4.8 THz.} \label{fig5}
\end{center}
\end{figure}

\begin{figure} 
\begin{center}
\vskip 0cm
\resizebox{1\linewidth}{!}{\includegraphics{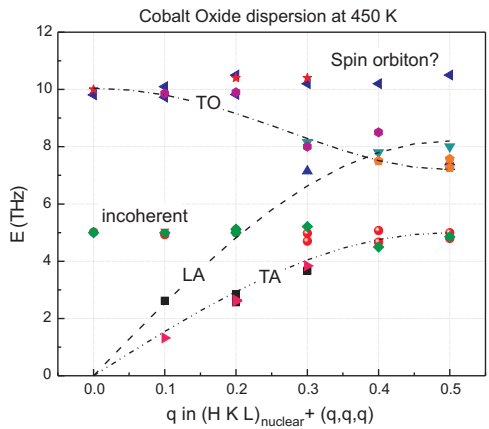}}\vskip 0cm \caption []{The high-temperature dispersion relations for excitations in cobalt oxide. The lines describe the expected phonons of the transverse acoustic, longitudinal acoustic and transverse optic branches. The branch of constant frequency, E, is understood to arise from the incoherent scattering from cobalt nuclei whose density of vibrational states peaks at $\sim$5 THz. The constant frequency branch at $\sim$10 THz is believed to be the spin orbital transition from the ground double to the lowest spin-orbit state. } \label{fig6}
\end{center}
\end{figure}

To separate the phonon from the $j$=1/2 to $j$=3/2 spin-orbit transition (orbiton) and spin excitations, inelastic scans were also performed at 450 K well above the N\'{e}el transition temperature. The high temperature scans exhibit well-defined phonons belonging to the transverse and longitudinal acoustic modes (TA and LA) and the transverse optic (TO) mode. An example in Fig.~\ref{fig5} shows the coherent LA phonon at 7.2 THz and the peak from the incoherent cobalt vibrational density of states at 4.8 THz, reduced slightly from its low temperature value ~\cite{sakurai68}. The dispersion observed for these modes is shown in Fig.~\ref{fig6}. The phonon frequencies are in reasonable agreement with the early phonon dispersion measurements~\cite{sakurai68} allowing for anharmonic effects at high temperatures.

The importance of the dispersion, however, lies in the fact that one branch of excitations at $\sim$10.2 THz in Fig.~\ref{fig6}, does not correspond to any phonon. Hence it may be attributed to a spin orbiton excitation from the $j$=1/2 ground state to the $j$=3/2 excited state, in which the coupled Co$^{2+}$ spin and orbital moments both change. Since the excited state is fourfold degenerate it is likely that the orbiton transition that remains well defined at large temperatures is the one from the ground state to that state whose transition energy is most weakly dependent on the exchange field. This is because in the paramagnetic state at 450 K there are large fluctuations about zero of the local exchange field that will broaden and render unobservable all other ground state transitions that depend strongly on local exchange field. An orbital transition at 10.2 THz and the Land\'{e} interval rule imply that the spin-orbit coupling parameter is 4.53 THz, reduced to 84$\%$ of its free-atom value by covalency.

Magnetic excitations of CoO have also been studied with optical spectroscopy~\cite{kant08,austin70}. A comparison of the frequencies observed with this technique to the frequencies observed with neutron scattering can provide further evidence for the origin of the observed peaks. Optical spectroscopy measurements have indicated that well into the AF phase, there are three magnetic excitations at 6.63, 7.50 and 8.84 THz. The frequencies of the excitations observed at 6.63 and 7.50 THz coincide with the two of the peaks we have observed. The high frequency excitation observed at 8.84 THz with optical spectroscopy has a different frequency from that observed with neutron scattering. We consider that this is confirmation that this mode arise from the zone boundaries of the magnetic zone from domains whose ordering wave vectors are not (0.5 0.5 0.5).

We have developed a model for the cobalt spin and orbital excitations that enables us to identify which is the exchange-insensitive spin-orbit transition. The high-spin and high-angular momentum of the Hund's rule model yields atomic states that couple together to give a band of magnetic dipole excitations that carry both spin and orbital weight. A second inference within the Hund's rule model is that the spin orbiton becomes, in the ordered antiferromagnetic state, a transition of symmetry, S$^+$ (S$^-$), that is, it is a spin excitation that raises (lowers) the spin on a spin-up (spin-down) site rather than a conventional spin wave that turns an aligned spin down.

With an ordering wavevector of the form $<$0.5 0.5 0.5$>$ there are four possible spin ordering domains. Within each domain there are several bands of spin excitations to states controlled by the exchange, spin-orbit and crystal field. Our numerical calculations of the magnetic modes within the Hund's rule model show that the two out-of-plane domains have the same frequency so that the four spin domains then give rise to only three frequencies for each ground-state transition. With only two transitions, the spin wave and the lower orbiton transitions in play, we expect six frequencies at each wavevector transfer \textbf{Q}. Our calculation leads to a model that has the same structure as in Refs.~\cite{sakurai68,buyers71} and adopted by~\cite{tomiyasu06} but with parameters that are different from both earlier references~\cite{sakurai68,tomiyasu06}. Details of the theoretical model will be published elsewhere.

\section{Conclusion}
We have studied the magnetic, orbiton and phonon excitations in orbitally active CoO. Our measurements at the magnetic zone centres indicate that there is a large anisotropy gap exceeding 4 THz. This gap arises from large exchange-induced mixing of the $j$=3/2 spin-orbit states into the $j$=1/2 ground doublet. Between 4 and 14 THz, we find at least four peaks. From measurements as a function of Q and temperature, we have identified three largely magnetic excitations at 6.5, 7.6 and 9.5 THz. The peak at 4.9 THz also has some magnetic weight but it largely carries phonon amplitude. A possible spin orbiton excitation at 10.2 THz in the paramagnetic phase has been identified.

\section{Acknowledgements}
We are grateful to R. Sammon, T. Whan, R. Donaberger, J. Fox and L. McEwan at CNBC, Chalk River Laboratories for their excellent technical support during the experiments.

\end{document}